# Resonant Transmission Through Double Quantum Dots: Model and Numerical Results

Yuhui He
Department of Microelectronics, Peking University, Beijng 100871, China
Qin-Wei Shi
Department of Physics, University of Science & Technology of China, Hefei 230026 China

**Abstract:** Via projection operator technology, we restrict our discussion of Double Quantum Dots System in subspaces of fixed electron population. When an electron from outside tries to pass through the dots, we find transmission peaks occur, if the energy of the whole system scans through eigen-levels in one-more-electron subspace of the dots. We also discover selection rules for this resonant transmission. We attribute some of these rules to the differences between many-electron bonding states and anti-bonding states of the dots.

## INTRODUCTION

Nowadays, there have been numerical theoretic studies of transmission through quantum dots [1 2 3 4]. Among these studies, S-J Xiong and Y Xiong [1] restricted the system containing single quantum dot (QD) in subspaces of fixed electron population. Assuming that there is no more than one electron in the leads, they carried out an exact calculation of electron transmission through single QD, in the Coulomb blockade regime. R. Swirkowicz et al [2] treated the Coulomb interaction term based on Non-equilibrium Green's Function technology. By counting the Coulomb interaction term into self-energy and using Hartree-Fork approximation [5], they decoupled higher order correlations of Green's Function. Then transmission properties are calculated from the famous stationary current formula by Jauho et al[6].

Among subjects in this field, we pay special attention to Double Quantum Dots (DQD) system[789]. There are several reasons. Firstly, DQD is a most ideal model for hydrogen molecular. As is known, there have been long-drawn efforts to understand relations between Molecular Orbital method and Valence Bond method[10]. By investigating into electron transport properties of DD, we may discover something interesting about the internal electronic structures of this "artificial molecular". Secondly, we can apply two gate voltages on the two dots separately. Thus we could tune onsite energies of the two dots independently. In this way, we can also imitate a molecular combined by two different atoms.

In this work, we study electronic transport through Double Quantum Dots connected to left lead and right lead. We restrict the energy range of the total system in three electron subspace, and also allow no more than one propagating electron in the leads. Then, with projection operator technology, we diagonalize the dots Hamiltonian, which has two terms of intra-dot Coulomb interactions. After diagonalization of the dot Hamiltonian, we get two-electron and three-electron eigen-states in population subspaces of the dots. Then, we continue our discussion of electron transmission in this new representation. To apply to Kondo effect regime, we make further assumption that in the interested energy range, either dot has only two spin-degenerate single-electron energy levels.

Although the model we proposed here is just a toy one, we think we get some interesting findings, include many-electron bonding states and anti-bonding states of the dots. The most interesting finding is the "Selection Rules", which decide whether the dots states could transit from bonding states (or anti-bonding states) in one subspace to bonding-states (or anti-bonding states) in another subspace, by adding or removing an electron on the dots.

Finally, based on our results, we explore the possibility of DQD working as quantum computing devices.

## Model

We consider two quantum dots sit adjacently, with one-dimensional wires connected to contacts. The system Hamiltonian consists of three components

$$\widehat{H} = \widehat{H}_L + \widehat{H}_D + \widehat{H}_T \qquad (1.1)$$

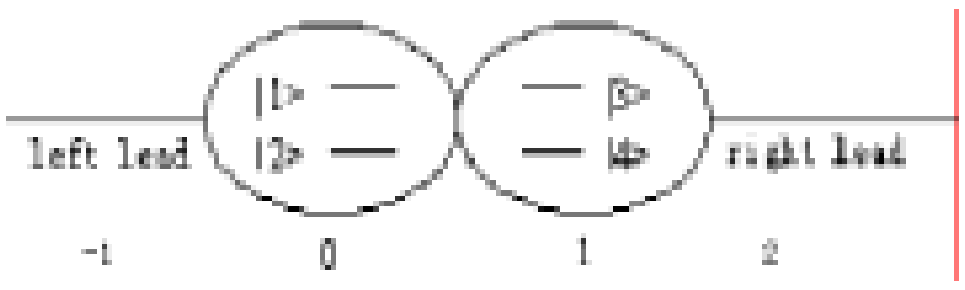

**Fig.1** **Sketch of the system**

And we apply tight-binding approximation to the lead Hamiltonian

$$\hat{H}_L = \sum_{m\neq -1,0,1} t_0(\hat{a}_m^+\hat{a}_{m+1} + c.c) + \sum_{m\neq 0,1} \varepsilon_0 \hat{a}_m^+\hat{a}_m \qquad (1.2)$$

$\hat{a}_m^+$ and $\hat{a}_m$ are onsite creation-annihilation operators in the leads.

We assume there are two single-electron energy levels in either dot. Energy levels 1 and 2 lie in the left dot, while 3 and 4 lie in the right dot. Other energy levels are so far away from the energy range of interest that, it is reasonable to neglect their influences

$$\hat{H}_D = \sum_{i=1}^{4} V_i \hat{c}_i^+\hat{c}_i + \sum_{\substack{i=1,2;\\ j=3,4}} (V_{ij}\hat{c}_i^+\hat{c}_j + c.c) + \frac{e^2}{2C}(\sum_{i=1,2}\hat{c}_i^+\hat{c}_i)^2 + \frac{e^2}{2C}(\sum_{j=3,4}\hat{c}_j^+\hat{c}_j)^2 \qquad (1.3)$$

In the above expression, $\hat{c}_{1(2)}^+$, and $\hat{c}_{1(2)}$ are creation-annihilation operators of the left dot on site 0, and $\hat{c}_{3(4)}^+$ and $\hat{c}_{3(4)}$ are the ones of the right dot on sight 1. $\{|1>, |2>\}$ are single-electron eigen-states for the left dot, while $\{|3>, |4>\}$ are that for the right dot. We can see the first term on the right hand of (1.3) represents the eigen-energies of the dots, which can be modulated by external voltages. The second term is the coupling of the two dots. The last two terms are Anderson-type models for Coulomb interactions in the two dots. It is known that Coulomb interaction is hard to deal with, since it involves four creation-annihilation operators. We can also write down the couplings of leads with the dots

$$\hat{H}_T = T_1^L\hat{a}_{-1}^+\hat{c}_1 + T_2^L\hat{a}_{-1}^+\hat{c}_2 + T_3^R\hat{a}_2^+\hat{c}_3 + T_4^R\hat{a}_2^+\hat{c}_4 + c.c. \qquad (1.4)$$

Now, we assume that in the energy range of interest, there are always three electrons in the total system, which is a subspace of the total system. Furthermore, there is always no more than one electron in the leads. The key of latter approximation is that, for the leads, we consider only one electron near the Fermion surface. With such approximation, the eigen-state wave function can be written as

$$|\varphi> = \sum_{m\neq 0,1; i>j} r_{m,i,j}\hat{a}_m^+\hat{c}_i^+\hat{c}_j^+|0> + \sum_{i>j>k} s_{ijk}\hat{c}_i^+\hat{c}_j^+\hat{c}_k^+|0> \qquad (1.5)$$

The first term on the right represents that there are two electrons on the dots, one electron in the leads. The second term represents three electrons are all on the dots, and we call it “interim state” or “virtue state”.

Considering Hamiltonian operates on this subspace, we can write down the annihilation operators of dots

$$\hat{c}_i = \sum_j |j>_A<i,j| + \sum_{j,k}|j,k>_A<i,j,k| \qquad (1.6)$$

The subscript ‘A’ means anti-symmetry law of Fermions. From now on, we neglect it for simplicity. And so are the creation operators

$$\hat{c}_i^+ = \sum_j |i,j>_A<j| + \sum_{j,k}|i,j,k>_A<j,k| \qquad (1.7)$$

Generally, the creation-annihilation operators can be expanded by projection operators as

$$\hat{c}_i^+ = |i><0| + \sum_j |i,j>_A<j| + \sum_{j,k}|i,j,k>_A<j,k| + \sum_{j,k,l}|i,j,k,l>_A<j,k,l| + \cdots \qquad (1.8)$$

$$\hat{c}_i = |0><i| + \sum_j |j>_A<i,j| + \sum_{j,k}|j,k>_A<i,j,k| + \sum_{j,k,l}|j,k,l>_A<i,j,k,l| + \cdots \qquad (1.9)$$

It is clear that to get (1.6) and (1.7), here we have made truncations on the expansion. The physical meaning of this approximation is that, influences of larger population subspaces have been neglected. With this approximation, we get some typical term for the Hamiltonian

$$\hat{c}_i^+\hat{c}_j = \sum_k |i,k><k,j| + \sum_{k,l}|i,k,l><j,k,l| \qquad (1.10)$$

And we find that, in this approximation, the Coulomb interaction terms in the dots Hamiltonian $H_D$ can be written as

$$\hat{H}_C = \frac{e^2}{2C}\{4|1,2><1,2| + 4|3,4><3,4| + 2\sum_{\substack{i=1,2;\\ j=3,4}}|i,j><i,j| + 5\sum_{i>j>k}|i,j,k><i,j,k|\} \qquad (1.11)$$

In this way, we have circumvented the Coulomb interaction terms.

Our purpose is to diagonalize the dot Hamiltonian, and get new basis of representation. And after some tedious treatment, we get the new basis for the dot Hamiltonian

$$\{|II,i>, \quad |III,j>\} \quad (i = 1,\cdots,6;\ j = 7,\cdots,10) \qquad (1.12)$$

In this new representation, $\widehat{H}_D$ is diagonalized

$$\widehat{H}_D=\sum_{i=1}^{6}V_{II,i}\,|\,II,i><II,i\,|+\sum_{j=7}^{10}V_{III,j}\,|\,III,j><III,j\,| \quad (1.13)$$

$II$ means that there are two electrons in the dots, and there are 6 such eigen-vectors. $III$ means that there are 3 electrons in the dots, and there are 4 such eigen-vectors. A little tip is worth mentioning here: when we write out $\widehat{H}_D$ in the original representation, only matrix elements in separate subspaces of $\{i,j\}$ or $\{i,j,k\}$ are non-zero. The crossing terms like $<l,m\,|\,\widehat{H}\,|\,i,j,k>$ are zero. So after diagonalization, $\{|\,II,l>\}$ involve $\{|\,i,j>\}$ only, while $\{|\,III,m>\}$ involve $\{i,j,k\}$ only. From diagonalization of dot Hamiltonian, we also get the unitary matrix ($U^{-1}\widehat{H}_D U=\widehat{H}_D^{'}$). With this unitary matrix, we can write down coupling Hamiltonian $\widehat{H}_T$ in this new representation as

$$\widehat{H}_T=\sum_{m=1}^{6}\sum_{n=7}^{10}(t_{-1,m,n}\widehat{a}_{-1}^{+}\,|\,II,m><III,n\,| \quad (1.14)$$
$$+t_{2,m,n}\widehat{a}_{2}^{+}\,|\,II,m><III,n\,|+c.c.)$$

Now, we write down eigen-states for the system in this new representation

$$|\,\varphi>=\sum_{m\neq 0,1}\sum_{i=1}^{6}P_{m,i}\widehat{a}_m^{+}\,|\,II,i>+\sum_{j=7}^{10}Q_j\,|\,III,j> \quad (1.15)$$

And we solve Schodinger Equation $\widehat{H}\,|\,\varphi>=E\,|\,\varphi>$, by matching coefficients of different eigen-vectors.

There are four kinds of equations

$$1)\widehat{a}_m^{+}\,|\,II,l> \qquad (m\neq -1,0,1,2;\quad l=1,\cdots,6)$$
$$t_0(P_{m+1,l}+P_{m-1,l})=(E-\varepsilon_0-V_{II,l})P_{m.l} \quad (1.16)$$

$$2)\widehat{a}_{-1}^{+}\,|\,II,l> \qquad (l=1,\cdots,6)$$
$$t_0P_{-2,l}+\sum_{n=7}^{10}t_{-1,l,n}Q_n=(E-\varepsilon_0-V_{II,l})P_{-1,l} \quad (1.17)$$

$$3)\widehat{a}_{2}^{+}\,|\,II,l> \qquad (l=1,\cdots,6)$$
$$t_0P_{3,l}+\sum_{n=7}^{10}t_{2,l,n}Q_n=(E-\varepsilon_0-V_{II,l})P_{2,l} \quad (1.18)$$

$$4)\,|\,III,n> \qquad (n=7,8,9,10)$$
$$\sum_{m=1}^{6}(t_{-1,m,n}^{*}P_{-1,m}+t_{2,m,n}^{*}P_{2,m})=(E-V_{III,n})Q_n \quad (1.19)$$

In this way, that every occupied state $|\,II,l>$ with one electron in the leads represents one channel. It is now a multiple-channel transmission case as in Fig.2.

Now we assume that there is one electron coming from the left lead, in Channel 1

$$P_{m,1}=\begin{cases}e^{ik_1(m+1)}+r_1e^{-ik_1(m+1)} & (m\leq -1)\\ t_1e^{ik_1(m-2)} & (m\geq 2)\end{cases} \quad (1.20)$$

In the multiple-channel framework, there will be reflecting and transmitting waves in all other channels.

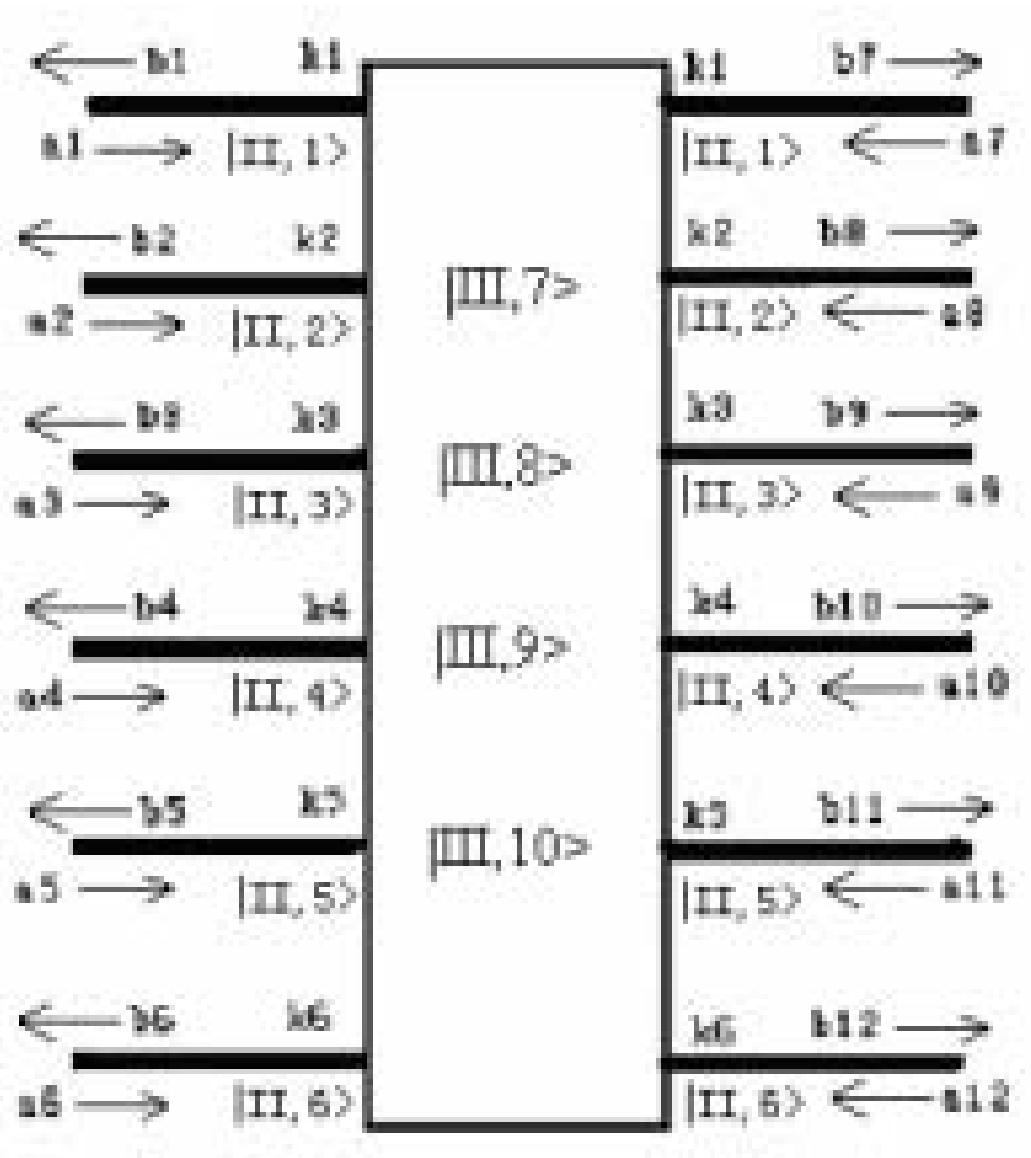

**Fig.2 Here $a_i$ represents incident wave, $b_i$ represents outgoing wave. For a given value of system energy E, different dots energies of $|\,II,l>$ will lead to different wave numbers of the incident electron, like $k_1$, $k_2$, etc.**

$$P_{m,j}=\begin{cases}r_je^{-ik_j(m+1)} & m\leq -1\\ t_je^{ik_j(m-2)} & m\geq 2\end{cases} \quad (1.21)$$
$$(j=2,3,4,5,6)$$

With this approach, Equations (1.14-17) get further simplified. First, we get the $E(k_j)$ relation for the system

$$2t_0\cos k_j+V_{II,j}+\varepsilon_0=E \qquad (j=1,\cdot\cdot,6) \quad (1.22)$$

This formula is as expected. $V_{II,j}$ is the eigen-energy of two electrons on the dots, while $(2t_0\cos k_j+\varepsilon_0)$ is the familiar term for single electron in the lead. We also get other 16 equations for 16 unknown coefficients ($r_{i=1,\cdot\cdot,6}$, $t_{i=1,\cdots,6}$, $Q_{7,8,9,10}$). Solving these equations, we get 12 elements for the first column of the S-matrix

$$\begin{bmatrix} b_1 \\ b_2 \\ \vdots \\ b_6 \\ b_7 \\ \vdots \\ b_{11} \\ b_{12} \end{bmatrix} = \begin{bmatrix} r_1 & & \\ r_2 & & \\ \vdots & & \\ r_6 & & \\ t_1 & & \\ \vdots & & \\ t_5 & & \\ t_6 & & \end{bmatrix} \begin{bmatrix} a_1 \\ a_2 \\ \vdots \\ a_6 \\ a_7 \\ \vdots \\ a_{11} \\ a_{12} \end{bmatrix} \qquad (1.23)$$

Similar approaches can be applied to other 11 channels to get the whole S-matrix. From Landauer Type formula[11], we can calculate transmission coefficients and Conductance in ballistic transport regime.

# Numerical Results and discussion

Now, let's apply our model to cases near Kondo effect regime. We assume that the two energy levels in separate dots are in fact spin-degenerate levels. That is, $|1>$ and $|3>$ are for spin-up, while $|2>$ and $|4>$ are for spin-down, as in Fig.3. Thus, it is reasonable to set the coupling strength of $|1>$ with $|4>$ to be zero, and so is $|2>$ with $|3>$. Also, we set $V_1 = V_2 = V_3 = V_4$. So we can see the four single-electron states are degenerate.

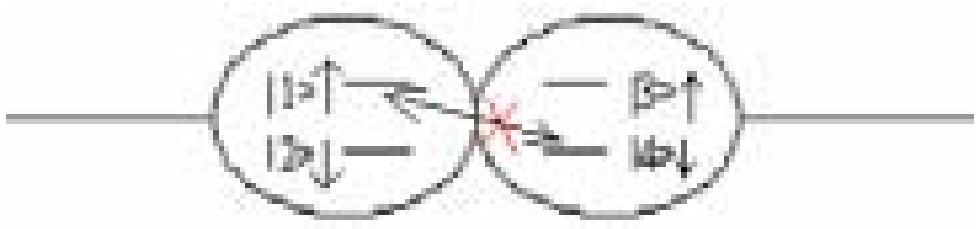

**Fig.3 spin-flip coupling of the dots are forbidden in this model for simplicity.**

Generally, there are two limits that we should pay separate attention. First, the life time of excited states in the dots are so short that, only the ground state of the quantum dots is occupied. In this case, after an electron from outside passes through (or reflects back) the dots, the two electrons on the dots will soon fall back to their ground state. Thus, when the next electron comes, it will see "the same" dots state as the former one. In this limit, there is no correlation between fore-and-aft electrons. Thus, we only need to consider transmission involving the lowest channel, as in Fig.4 and Fig. 5. The other limit is that, the excited states ($|II,l>(l=1,\cdots,5)$) have sufficient long live time. In this case, there will be correlations in the system. If an electron comes in one channel, and reflects (or pass) into another channel, the two electrons on dots will transit to another eigen-state. Since they can stay sufficient long time on the new state, a next electron from the lead will see "different" channel from the last one. In this way, the fore-and-aft incident electrons are correlated. Result is present in Fig.6.

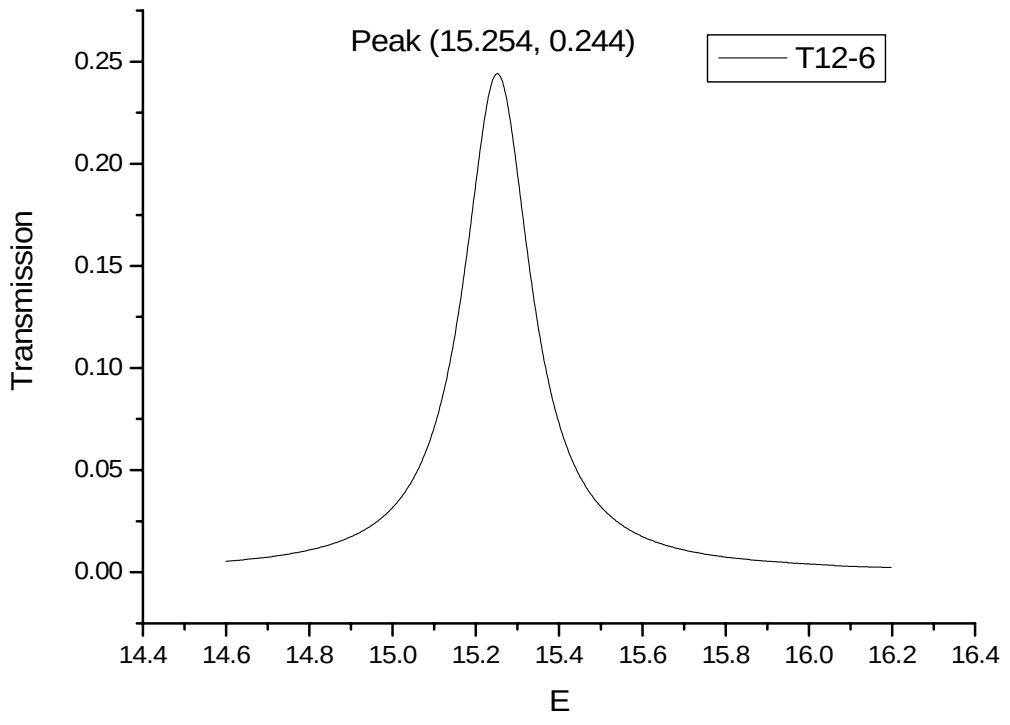

**Fig.4 Transmission from channel 6 to channel 12. This is for the short lifetime excited states limit.**

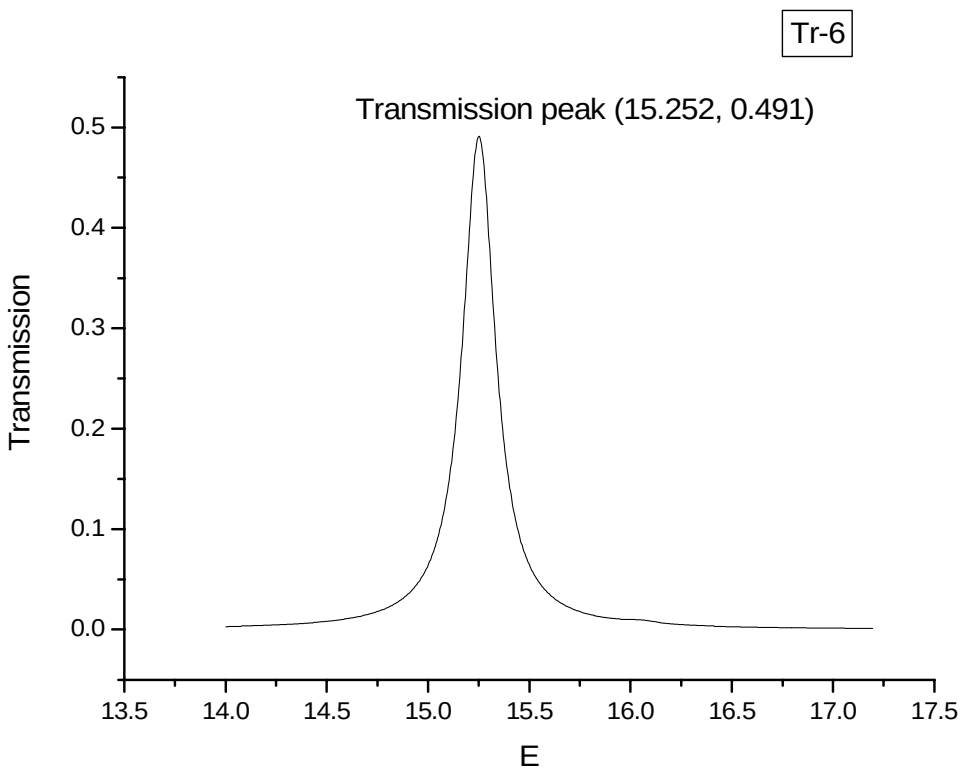

**Fig.5 Transmission from the lowest channel 6 to all channels at right-hand ( $T_{r\leftarrow 6} = \sum_{i=7}^{12} T_{i\leftarrow 6}$ ).This is for the short lifetime excited states limit.**

Parameters are set as $\varepsilon_0 = 0$, $T_{L(R)} = 0.05t_0$, $V_i = 0.5t_0$, $e^2/C = 0.025t_0$, $V_{13} = V_{24} = 0.04t_0$ and $V_{14} = V_{23} = 0$. For setting these parameters, we take S-J Xiong's work [1] as reference. The eigen-energies of the dots are as follows

$V_{\mathrm{II},1} = 11.185$

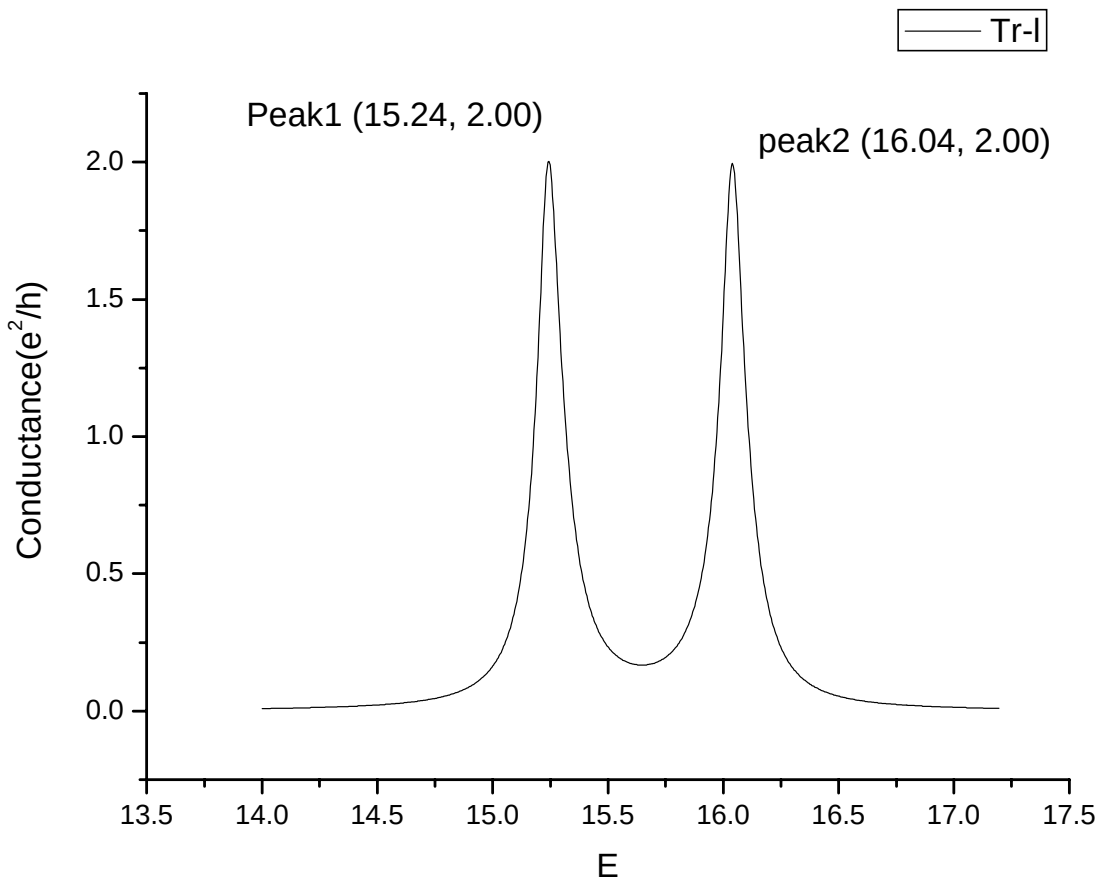

**Fig.6 Transmission from all left channels to all right channels ($T_{r\leftarrow l}=\sum_{i=7}^{12}\sum_{j=1}^{6}T_{i\leftarrow j}$). This is for long lifetime excited states limit.**

$$V_{II,2}=10.500$$
$$V_{II,3}=V_{II,4}=V_{II,5}=10.250$$
$$V_{II,6}=9.565$$
$$V_{III,7}=V_{III,8}=16.025$$
$$V_{III,9}=V_{III,10}=15.225$$

Here is interesting to notice the partial unraveling of the degenerate energy levels. It is easy to check the follows

$$|III,7>=\frac{1}{\sqrt{2}}(|1,2,3>+|1,3,4>)$$
$$|III,8>=\frac{1}{\sqrt{2}}(|1,2,4>+|2,3,4>)$$
$$|III,9>=\frac{1}{\sqrt{2}}(-|1,2,3>+|1,3,4>)$$
$$|III,10>=\frac{1}{\sqrt{2}}(|1,2,4>-|2,3,4>)$$

What distinguishes $|III,7(8)>$ from $|III,9(10)>$ are anti-bonding states and bonding states. Thus, their degeneration is unraveled. However, why are $|III,7>$ and $|III,8>$ still degenerate? Remember that we have set $V_{13}=V_{24}=0$, while $V_{14}=V_{23}=0$.For this kind of symmetry, some levels are still degenerate. For the same reason, since $|II,1>$ and $|II,6>$ are bonding state and anti-bonding state, their degeneracy is unraveled. On the other hand, $V_{II,3}$, $V_{II,4}$ and $V_{II,5}$ are degenerate because of the remained symmetry of the system.

$$|II,1>=\alpha(|1,2>+|3,4>)+\beta(|1,4>+|2,3>)$$
$$|\alpha|^2+|\beta|^2=1/2$$
$$|II,2>=\frac{1}{\sqrt{2}}(-|1,2>+|3,4>)$$
$$|II,3>=|2,4>$$
$$|II,4>=|1,3>$$
$$|II,5>=\frac{1}{\sqrt{2}}(-|1,4>+|2,3>)$$
$$|II,6>=-\beta(|1,2>+|3,4>)+\alpha(|1,4>+|2,3>)$$

Looking into Fig.4 and Fig.5, we can see very clear that, the transmission peak takes place, when the system energy scans through $|III,9(10)>$, the lower energy levels of three-electron eigen-states of the dots. This is typical resonant transmission phenomenon. However, it is puzzling why there is no transmission peak when energy scans through $|III,7(8)>$, the higher energy levels of three-electron eigen-states? The answer can be attributed to "selection rules". Let's look into the coupling Hamiltonian $\hat{H}_T$ in the new representation. Notice that $t_{-1,6,7}=0$ and $t_{-1,6,8}=0$. That is, the way of adding or removing a third electron on the dots, like $|II,6><III,7|$ and $|II,6><III,8|$, is forbidden. So there is no transmission peak near the eigen-energy of $|III,7(8)>$. We can also check that $t_{-1,1,9}=0$ and $t_{-1,1,10}=0$. That is to say, transitions of $|II,1><III,9|$ and $|II,1><III,10|$ are also forbidden. Thus, we allege that, if we calculate transmission from the highest channel involving dots state $|II,1>$, there will be no transmission peak near the eigen-energy of $|III,9(10)>$. It is confirmed in Fig.7.

Let's look into "Selection Rules" for details

$$|III,10>\not\rightleftarrows|II,1>\not\rightleftarrows|III,9>$$
$$|III,9>\not\rightleftarrows|II,3>\not\rightleftarrows|III,7>$$
$$|III,10>\not\rightleftarrows|II,4>\not\rightleftarrows|III,8>$$
$$|III,8>\not\rightleftarrows|II,6>\not\rightleftarrows|III,7>$$

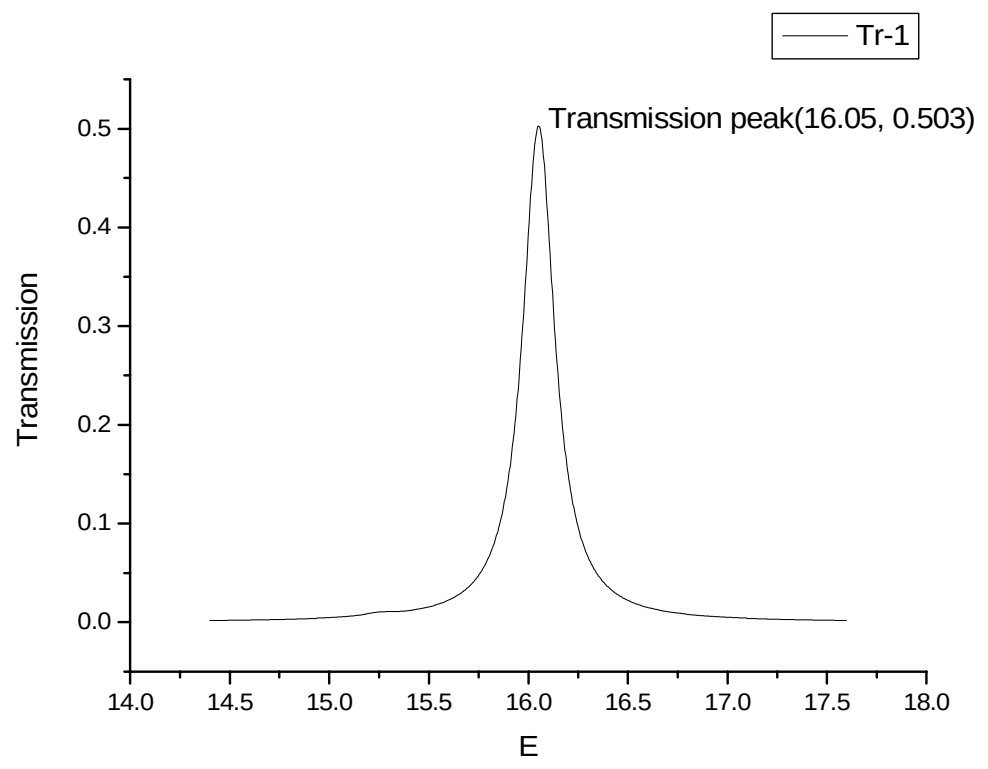

**Fig.7 Transmission from the highest channel involving $|II,1>$ ($T_{r\leftarrow 1}=\sum_{i=7}^{12}T_{i\leftarrow 1}$).There is a transmission peak near energy level of $|III,7(8)>$, but no peak appears near energy level of $|III,9(10)>$.**

Why do these "selection rules" exist? Among these rules, the 2nd and the 3rd are easy to understand. For example, since $|II,3>$ involves the state $|2,4>$ only, when we create an electron on this state, it is impossible to get the final states like $|1,2,3>$ or $|1,3,4>$. However, these two three-electron states are necessary components of $|III,7(9)>$. So it is now clear why transitions from $|II,3>$ to $|III,7(9)>$ are forbidden. To explain the 1st and 4th rule, let's take $|II,1><III,9|$ as example. It not only has components of $\frac{\alpha}{\sqrt{2}}(|3,4><1,3,4|-|1,2><1,2,3|)$ and $\frac{\beta}{\sqrt{2}}(|1,4><1,3,4|-|2,3><1,2,3|)$, but also have components of $\frac{\alpha}{\sqrt{2}}(|1,2><1,3,4|-|3,4><1,2,3|)$ and $\frac{\beta}{\sqrt{2}}(|2,3><1,2,4|-|1,4><2,3,4|)$. Progresses represented by the latter two components are impossible. Careful calculation shows it is this requirement that makes $t_{-1,1,9}=0$, thus forbidding progress of $|II,1><III,9|$. However, since other transition progresses also have some forbidden components (For example, $|II,1><III,7|$ have components of $\frac{\alpha}{\sqrt{2}}(|1,2><1,3,4|+|3,4><1,2,3|)$ and $\frac{\beta}{\sqrt{2}}(|1,4><1,3,4|+|2,3><1,2,3|)$ ), why are they not forbidden? The answer lies in the differences between $|III,7>$ and $|III,9>$. $|III,7>$ is an anti-bonding state, while $|III,9>$ is the corresponding bonding states. The symmetry of them is opposite. The same things happen to all other bonding states and corresponding anti-bonding states in 1st and 4th rules.

We may summarize two selection rules for adding or removing an additional electron on the dots.

**1)** It is impossible to change more than one single-electron-state component of a many-electron dots state by adding or removing an additional electron on the dots.

**2)** Transmissions are forbidden between anti-bonding (bonding) states in lower subspace of the dots and bonding (anti-bonding) states in higher subspace.

We can check our selection rules for the case of two electrons in system. Following similar approach, we diagonalize Dot Hamiltonian $H_D$ in the subspace of one-electron occupation, and get

$$|I,1>=\frac{1}{\sqrt{2}}(|1>+|2>)$$

$$|I,2>=\frac{1}{\sqrt{2}}(|2>+|4>)$$

$$|I,3>=\frac{1}{\sqrt{2}}(-|1>+|3>)$$

$$|I,4>=\frac{1}{\sqrt{2}}(|2>-|4>)$$

And the eigen-energies are $V_{I,1}=V_{I,2}=5.525$ and $V_{I,3}=V_{I,4}=4.725$. It is clear that $|I,1>$ and $|I,2>$ are anti-bonding states, while $|I,3>$ and $|I,4>$ are corresponding anti-bonding states. We can write out "selection rules" based on the above equations directly, or check out them by numerical results

$$\begin{cases} |I,1> \not\rightleftharpoons |II,3> \\ |I,2> \not\rightleftharpoons |II,4> \\ |I,3> \not\rightleftharpoons |II,3> \\ |I,4> \not\rightleftharpoons |II,4> \end{cases}$$

$$\begin{cases} |I,1> \nLeftrightarrow |II,6> \\ |I,2> \nLeftrightarrow |II,6> \\ |I,3> \nLeftrightarrow |II,1> \\ |I,4> \nLeftrightarrow |II,1> \end{cases}$$

The first group is decided by Rule 1, and the second group is by 2. Numerical results for this two-electron system are represented in Fig.8 and Fig.9. Fig.8 is for short lifetime excited states limit, while Fig.9 is for long lifetime excited states limit.

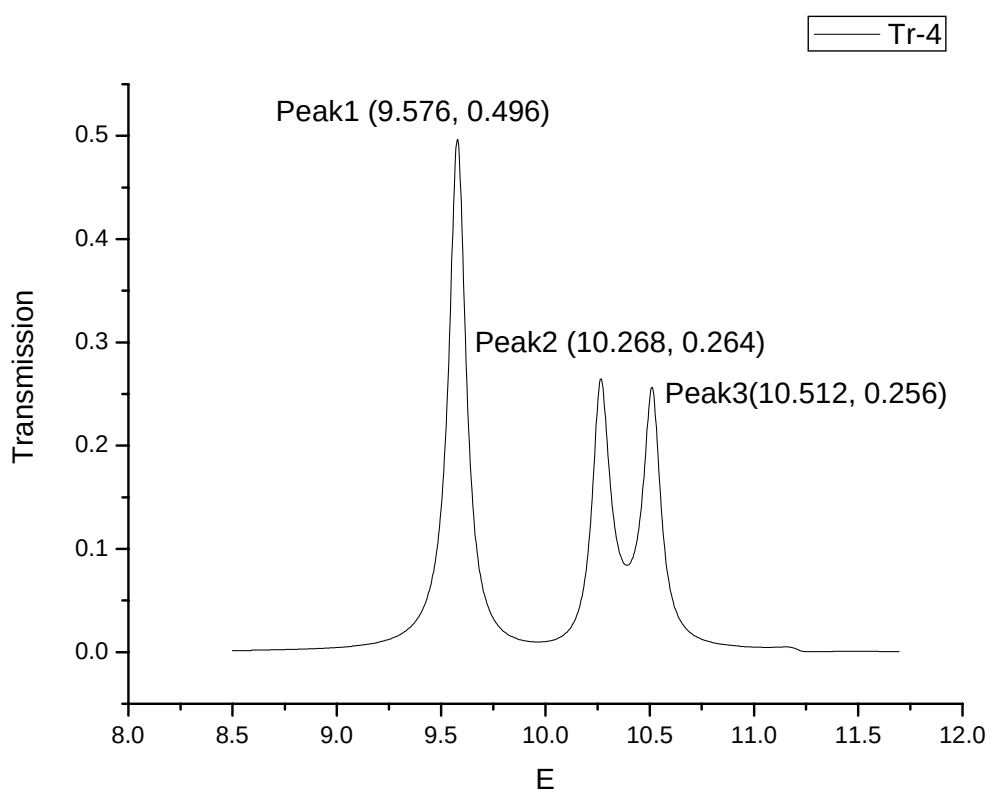

**Fig.8 Transmission from the lowest channel involving dots eigen-state $|I,4>$ ($T_{r\leftarrow 4}=\sum_{i=5}^{8} T_{i\leftarrow 4}$).There are 3 peaks near $V_{II,2}$, $V_{II,3}=V_{II,4}=V_{II,5}$ and $V_{II,6}$. However, as expected, there is no transmission peak near $V_{II,1}$. This is for short lifetime excited states limit.**

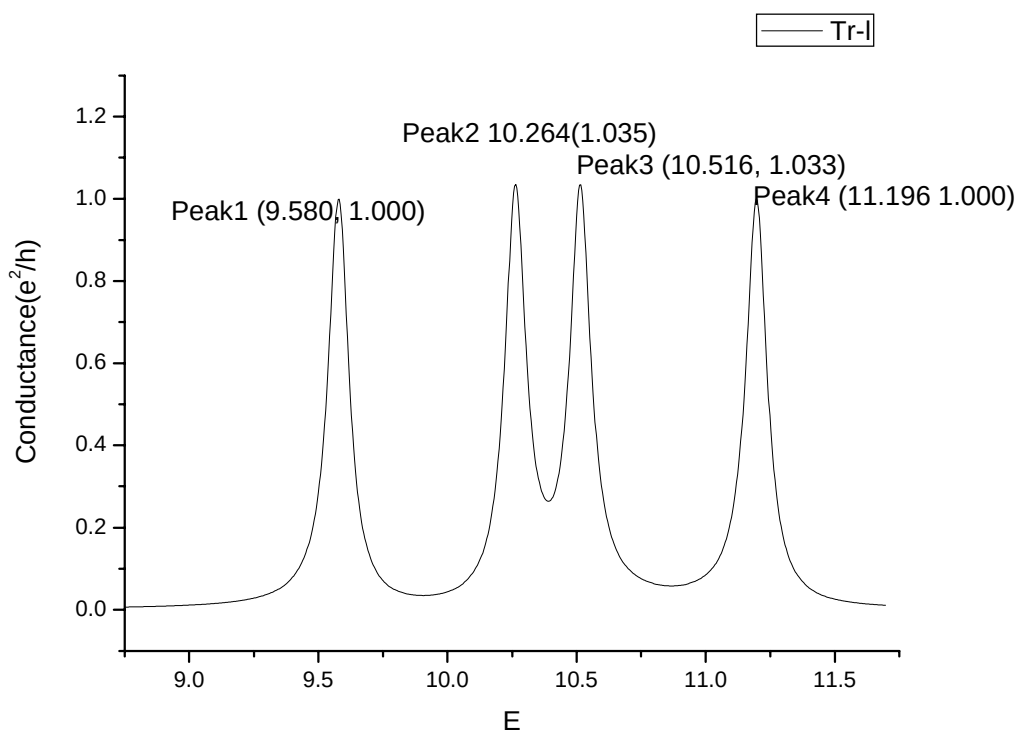

**Fig.9 Transmission from all left channels to all right channels ($T_{r\leftarrow l}=\sum_{j=5}^{8}\sum_{i=1}^{4} T_{j\leftarrow i}$). There are four peaks near all possible two electron eigen-energy levels of the dots. This is for long lifetime excited states limit.**

## Summary

In this work, we discussed transmission properties of Double Quantum Dots. Via projection operator technology, we restricted the system in fixed population subspace. We applied our model to Kondo effect regime, and found the virtual physics progress for transmission: an electron comes from the lead, and jumps into the Dots (if it could). Thus the dots states transfer to higher subspace consisting of one-more electron. Once the additional electron jumps out, the dots states will fall back to original population subspace. What we found interesting are “Selection Rules”, which seem to govern the “jump-into” and “jump-out” progresses. Among these two rules, the one forbidding transitions between many-electron anti-bonding states and bonding states in different subspaces is most fascinating, and needs further investigation.

We suggest that, further research may include mechanisms for excited states relaxation, in an easy-dealing way. Thus by investigating into dynamic equations, we may turn the virtual progresses to real ones in time domain.

For application purpose, we think that the limit of long-lifetime excited states may be used for potential executing devices of quantum bits. The principle is that after one electron passes through the dots, the dots state will change in its original subspace. Because of correlations between fore-and-aft passing electrons, this device will have some “memory effect”. Another obvious advantage is that we tune the states on the quantum dots by incident electrons, but not LASER or magnetic fields. This is crucial for integrated circuits.

## Acknowledgement

Yuhui He, a student who has not begun his graduate career, is financially supported by Prof. Jie Chen personally. Besides, we are grateful to Prof. Ming-Wei Wu for beneficial discussion. He also thanks graduate students Danqiong Hou and Hui Guo heartily for their help.

---

[1] Shi-Jie Xiong and Ye Xiong, Phys. Rev. Lett. **83**, 1407 (1999).
[2] R Swirkowicz, J Barnas, and M Wilczynski, Phys. Rev. B **68**, 195318 (2003).
[3] A A Clerk and P W Brower, Phys. Rev. Lett. **86**, 4036 (2001).

[4] P Orellana and F Claro, Phys. Rev. Lett. **90**, 178302 (2003).

[5] Hartmut Haug and Antti-Pekka Jauho, *Quantum Kinetics in Transport and Optics of Semiconductors*, **1st** edition (Springer-Verlag, Heidelberg, 1998), Vol. **1**, Chap. 12, p.173

[6] Antti-Pekka Jauho and N S Wingreen, Phys. Rev. B **50**, 5528 (1994).

[7] M L Ladron se Guevara, F Claro, and Pedro A Orellana, Phys. Rev. B **50**, 5528 (1994).

[8] Konstantin Kikoin and Yshai Avishai, Phys. Rev. Lett. **86**, 2090 (2001).

[9] J Fransson, Phys. Rev. B **69**, 201304 (2004).

[10] Roald Hoffman, Sason Shaik, and Philippe C Hiberty, Acc. Chem. Res. **36**, 750 (2003).

[11] R Landauer, Physica Scr. **T42**, 110 (1992).